**Title:**

Superresolution imaging of single DNA molecules using stochastic photoblinking of minor groove and intercalating dyes

**Author names and affiliations:**

Helen Miller, Zhaokun Zhou, Adam J. M. Wollman, Mark C. Leake[*]

Biological Physical Sciences Institute (BPSI), Departments of Physics and Biology, University of York, York, YO10 5DD

[*] Corresponding author.

*E-mail address*: mark.leake@york.ac.uk (M. C. Leake)

**Abbreviations**

DIG – digoxigenin, FWHM - full width at half maximum, λ DNA - Lambda phage DNA

PSF - point spread function, SNR - signal-to-noise ratio.

**Highlights:**

- A bifunctional DNA construct tethers single molecules to paramagnetic beads
- Intercalating (YOYO-1) and minor groove (SYTO-13) dyes are bound to DNA and photoblink
- Stochastic dye photoblinking is used to obtain superresolution information via localization microscopy

**Abstract:**

As proof-of-principle for generating superresolution structural information from DNA we applied a method of localization microscopy utilizing photoblinking comparing intercalating dye YOYO-1 against minor groove binding dye SYTO-13, using a bespoke multicolor single-molecule fluorescence microscope. We used a full-length ~49 kbp λ DNA construct possessing oligo inserts at either terminus allowing conjugation of digoxigenin and biotin at opposite ends for tethering to a glass coverslip surface and paramagnetic microsphere respectively. We observed stochastic DNA-bound dye photoactivity consistent with dye photoblinking as opposed to binding/unbinding events, evidenced through both discrete simulations and continuum kinetics analysis. We analyzed dye photoblinking images of immobilized DNA molecules using superresolution reconstruction software from two existing packages, rainSTORM and QuickPALM, and compared the results against our own novel home-written software called ADEMS code. ADEMS code generated lateral localization precision values of 30-40 nm and 60-70 nm for YOYO-1 and SYTO-13 respectively at video-rate sampling, similar to rainSTORM, running more slowly than rainSTORM and QuickPALM algorithms but having a complementary capability over both in generating automated centroid distribution and cluster analyses. Our imaging system allows us to observe dynamic topological changes to single molecules of DNA in real-time, such as rapid molecular snapping events. This will facilitate visualization of fluorescently-labeled DNA molecules conjugated to a magnetic bead in future experiments involving newly developed magneto-optical tweezers combined with superresolution microscopy.

**1. Introduction**

In the last decade, single-molecule methods of biological physics have generated enormous advances in our understanding of the 'living' component of what physicists describe as soft condensed matter [1,2]. There are now a wide range of experimental and analytical single-molecule tools available to the researcher [3,4]. Fluorescence microscopy is relatively non-invasive compared to many other biophysical techniques. Although some forms of superresolution microscopy suffer issues of phototoxicity, such as STED [5] with the high potential excitation laser intensities used, and conventional PALM/STORM [6,7] with their use of harmful ultraviolet laser activation wavelengths, many of the more recent forms of localization-based superresolution microscopy are minimally perturbative to the native cellular physiology [8]. The optical diffraction limit is circumvented by acquiring a series of images in each of which only a sparse subset of fluorophores are 'on' (i.e. photoactive), and numerically fitting the point spread function (PSF) pixel intensity profile of detected dye molecules from a high-sensitivity camera pixel array. Provided the individual PSFs are non-overlapping (i.e. the nearest-neighbor separation is greater than the standard optical resolution limit), the set of localization coordinates from the image stack can be combined into one reconstructed image with sub-pixel resolution.

DNA encodes the genetic information in all known organisms and is key to many cellular processes. The well-characterized double-helical structure makes DNA an extremely tractable molecule for single-molecule super-resolution studies – its structure has a spatial periodicity below the optical diffraction limit, and it has a well-defined sequence to which fluorescent probes can be conjugated via a range of robust chemical protocols. The periodic nature of the structure lends itself to coarse-graining and therefore simulations over physiologically relevant time scales. Lambda phage DNA ($\lambda$ DNA) is a double-stranded template of 48,502 base pairs with 12 base pair single stranded 'sticky ends' that allow for base-pairing type modifications at each end. This experimental flexibility has been utilized in many previous studies on cellular processes, such as exonuclease activity ([9]) and DNA repair ([10]).

One method to visualize DNA is to attach a fluorescent probe to it, for example by using a DNA-binding dye. Different types of DNA-binding dyes include intercalators, which fit between the base pairs of the helix, and minor groove binders, which attach between the helical backbones. A key advantage of such dyes is that their fluorescence emission intensity increases typically 100-1000 times upon binding to DNA [11], thus increasing the effective signal-to-noise ratio (SNR) for dye detection in the DNA sample. Intercalating dyes such as YOYO-1 have been well studied [12,13] but the minor groove binders, such as SYTO-13, have been less well studied despite being potentially less perturbative to the DNA structure [14].

Blinking assisted Localization Microscopy (BaLM) [15] is a superresolution method which utilizes the characteristic blinking and/or binding dynamics of fluorescent probes to achieve superresolution imaging by image subtraction. BaLM is one of a number of techniques (for example, see reference [16]) which do not require photoactivation or photoswitching, as in

PALM [6] and STORM [7] respectively. Binding-Activated Localization Microscopy (BALM) [17] utilizes the fluorescence enhancement for a dye when bound to nucleic acids compared to being free in solution. For the DNA binding dyes used in this work, photoblinking, rather than binding is the dominant mechanism (See section *3.4.2*), but the photoblinking is used to achieve separation of fluorophores to produce single molecule point-spread-functions, rather than to produce these by image subtraction.

Here we present the essential details of the key experimental assays for producing surface-immobilized and tethered DNA constructs for imaging using the photoblinking intercalating cyanine dye YOYO-1 and the minor groove binder SYTO-13. We model the binding/unbinding characteristics of these DNA binding dyes and compare three superresolution reconstruction packages on our data. We further show how a magnetic bead may be tethered to one end of the DNA construct to be used in future single-molecule superresolution fluorescence imaging experiments that will allow the DNA molecule to be structurally manipulated using magnetic tweezers.

## 2. Materials and Methods

*2.1 DNA construct design and preparation*

λ DNA purchased from NEB has single stranded 12 base, 'sticky-ends', which can be used to bind custom synthetic DNA by annealing phosphorylated complementary oligonucleotides and ligating (for example, see reference [18]). This enabled labeling the λ DNA with a biotin tag at one end and a digoxigenin (DIG) tag with Tex615 red fluorophore reporter probe at the other. Fig. 1a shows a schematic of the design, λ DNA in black and the synthetic ends in green and red. Table 1 contains the sequences of the synthetic oligonucleotides (IDT).

To label λ DNA, 10 μM synthetic DNA was first phosphorylated using 1 unit T4 polynucleotide kinase (Promega) in 1x T4 DNA ligase buffer (NEB) and incubating for 75 min at 37 $^0$C. The DIG end was annealed first in 1x T4 DNA ligase buffer with 1.5 nM λ DNA and 20 nM of oligos 1 and 2 at 65 $^0$C for 5 min. This mixture was allowed to cool back to room temperature before adding 1 unit of T4 DNA ligase (NEB) and incubating for 2 hours. To remove enzyme and excess synthetic DNA, a QIAEX II Gel Extraction Kit (Qiagen) was used. The biotinylated end was attached using the same protocol as for the DIG end with 20 nM oligos 3 and 4, before final clean up with the Gel Extraction Kit. The λ DNA construct was stored at 4 $^0$C or -20 $^0$C.

*2.2 Sample preparations*

Two sample preparation methods were developed to allow superresolution imaging. A simple, short preparation time assay which immobilizes DNA molecules to a glass coverslip, and a second assay for tethered DNA.

To achieve a suitable labeling density for superresolution imaging, dilutions of reagents were performed in 1x phosphate buffered saline (PBS; 10 mM phosphate buffer, 2.7 mM potassium chloride, 0.137 mM sodium chloride, pH 7.4, Sigma Aldrich). Two DNA dyes were utilized; the minor groove binder SYTO-13 (5 mM Solution in DMSO, Life Technologies Ltd.) diluted 1:99 in PBS to a concentration of 50 µM, and the intercalating YOYO-1 Iodide (1 mM Solution in DMSO, Life Technologies Ltd.) diluted 1:99 in PBS to a concentration of 10 µM.

Due to the large size of the λ DNA the effect of shearing forces was minimized by using large diameter pipette tips. For details of the DNA construct developed from λ DNA see section *2.1*.

Coverslips were plasma cleaned prior to use to remove impurities and reduce autofluorescence. Coverslips were subjected to a high frequency plasma for 1 minute in a Harrick PDC-32G plasma cleaner, handled with forceps, and were used immediately after treatment, since the effects of plasma cleaning relax exponentially with a time constant of around 4 hours [19].

*2.2.1 Immobilized DNA assay*

For the immobilization assay, samples of reagent were made up in 10µl volumes, composed of 5µl of the diluted dye to be used, and 5µl of the DNA sample (λ DNA was diluted 1:9 from stock to 1.6nM, DNA constructs were used at 0.1 nM). 5 µl of the sample was placed in the center of a glass slide. A plasma cleaned coverslip was gently dropped onto the liquid with a pair of forceps such that spreading was slow and bubble formation minimized. The chamber height confines the DNA to lie in a plane that is readily found during imaging. The edges of the sample were sealed with nail varnish to prevent evaporation. Samples were immediately transferred to the microscope for imaging.

*2.2.2 Tethered DNA assay*

To produce DNA tethers, tunnel slides were constructed by laying two lines of double sided tape 3mm apart on a standard microscope slide to produce tunnels of approximately 5µl volume [20]. Coverslips were dropped onto the tape and tapped down (avoiding the region above the tunnel which will be imaged) to ensure a good seal. Excess tape was removed with a razor blade. Slides were prepared via sequential flow steps after each of which was an incubation in an inverted position (coverslip down) for 5 minutes in a humidity chamber.

Flow cells were prepared with 5 µl of anti-digoxigenin (1 µg/ml, Roche Diagnostics) introduced to the tunnel slide by capillary action, producing a carpet of spatially separated antibody covering 5.5% of the surface. The remaining free surface was blocked by a 100 µl wash of Bovine Serum Albumin (BSA, 0.1 %, 1 mg/ml, Sigma Aldrich). Next, DNA/DNA construct (5 µl) was introduced to the channel, followed with a 100 µl PBS wash and then 5 µl of dye dilution. A further 100 µl wash with PBS was performed prior to imaging. This produced fluorescently-labeled DNA/DNA constructs that were not immobilized along their length, demonstrated movement under flow, and cleaved under prolonged laser illumination - presumably due to the action of free radicals produced by photobleaching.

A paramagnetic bead (Dynabeads® MyOne™ Streptavidin C1, Life Technologies Ltd.) can be conjugated to the DNA construct for use in magnetic tweezers experiments [21] by pre-incubation of 5µl 0.1nM DNA construct, 10 µl dye dilution and 5µl paramagnetic beads (100 µg/ ml), with mixing by manual finger flicking every 2.5 minutes for 30 minutes to avoid sedimentation.

*2.3 Fluorescence microscopy and imaging protocols*

We constructed a home-built microscope designed around a commercial Nikon Eclipse Ti-S inverted microscope body, with modified input excitation and output emission optics (Fig.1b). Input excitation was from a white-light supercontinuum laser (Fianium SC-400-4, Fianium Ltd.), filtered through a multi band pass excitation filter (Brightline 479-585, Semrock) to generate excitation centered on 479 nm and 585 nm for green and red excitation respectively. The beam generates a high intensity epifluorescence excitation field at the sample of full width at half maximum (FWHM) 24 µm and unattenuated intensity 1800 W/cm$^2$ and 2900 W/cm$^2$ for green and red excitation respectively via a 100X, NA 1.45 oil immersion objective lens (Nikon). Fluorescence emissions were imaged onto two EMCCD camera detectors (iXon Ultra 897, Andor Technology Ltd) giving 160 nm/pixel magnification via a spectral color splitter (TuCam, Andor Technology Ltd) allowing simultaneous imaging in two colors. The two color channels were registered relative to each other using 1 μm paramagnetic beads, which are visible in both channels. Data acquisitions were made at the full EM gain of absolute value ~300, using continuous video-rate sampling of 40 ms per frame, typically acquiring 1,000 frames per acquisition. Where appropriate, brightfield movies were acquired to determine the positions of bound paramagnetic beads.

*2.4 Image analysis*

We tested three methods to generate superresolution reconstructions of fluorescently labeled DNA, the open source QuickPALM [22] and rainSTORM software [23], specifically designed for reconstructing data of this type, and our own custom Matlab™ software designed originally for tracking diffusing molecules in single living cells [24,25] based on an Algorithm involving Dilation/Expansion in Matlab for Superresolution localization, or *ADEMS* code. All of these have similar functionality; they identify bright spots in each frame, quantify their intensity and size and assess - using different metrics - the 'quality' of each detected spot.

*2.4.1 QuickPalm*

QuickPALM finds bright spots in images using the Högbom 'CLEAN' method [26] and determines the spot centroid from its centre of mass. The code is Java based and runs as a plugin to ImageJ. It does not output localization precision metrics.

*2.4.2 rainSTORM*

The rainSTORM software segments images to find bright spots using a top-hat algorithm to even out the background and then thresholds the resultant image. The intensity centroid of these candidate spots is found using iterative Gaussian masking [27] and other parameters are fitted including the 2D PSF widths. Spots are rejected if they have too imprecise localization precision (We used >50nm, the default for this software) defined using Thompson's equation [27], based on the spot photon count compared to the background photon count. Each accepted spot is used to reconstruct the superresolution image.

*2.4.3 ADEMS code*

*ADEMS* code has similar functionality. It also uses a thresholded top-hat transformation but with the addition of a dilation/expansion step followed by an erosion to reduce spurious candidate spots. 2D Gaussian masking is used to generate an estimate for the intensity centroid, followed by a second stage involving a 2D Gaussian fit in which the Gaussian amplitude, local background offset and separate Gaussian sigma width values in *x* and *y* are free to vary but with the centroid coordinates from the first masking stage fixed. This novel two-stage approach resulted in greater robustness for fitting, converging at the very low values of SNR equivalent to dim single dye molecule signals, compared to using a fully unconstrained one-stage 2D Gaussian fit.  A useful output from the fitting algorithm is the integrated pixel intensity for each spot, which can be used in other imaged biological molecules to estimate molecular subunit stoichiometry of complexes [28, 29]. Different criteria are used for accepting a spot - the SNR, defined as the integrated spot intensity divided by the integrated background intensity over the same detected spot area, must be above a user-defined threshold value which can be correlated to robust statistical probabilistic confidence criteria. Localization precision can be calculated from the standard deviation of the intensity centroid position over time as well as independently by cluster analysis, which links proximal fluorescent spots together into a cluster and provides a measure of the localization precision from the mean intra-cluster distances. A copy of ADEMS code is available from the authors on request.

*2.5 Stochastic photoactivity analysis*

We analytically modeled the binding kinetics of the DNA-binding dyes with accepted values of binding constants to verify that the experimentally observed fluorescence emission events in our assay are mainly due to photoblinking rather than dynamic binding. We simulated the photoblinking of the dyes, linking fluorescence emissions to experimental parameters such as the laser excitation intensity, DNA and dye concentrations, and camera settings, so that the conclusions we draw from observations could be supported by realistic simulations.

*2.5.1 Binding and unbinding modeling of DNA dyes*

In principle, both photoblinking and transient binding could contribute to stochastic single fluorescence emission events which permit nanoscale localization of the fluorophores. YOYO-1 is a fluorogenic dye; its brightness increases by typically two orders of magnitude when bound to DNA compared to when free in solution [11].  To estimate the extent to which photoblinking or transient binding contribute to the measured localization, we calculate the on-rate at which free dye molecules bind to DNA in equilibrium, and the average binding lifetime that a dye stays bound to DNA;  a low on-rate indicates that fluorescence emission events are rarely due to new binding of free dye from solution, whereas a long binding lifetime in principle allows a typical dye molecule more time to stay on the DNA to undergo cycles of photoblinking.

The kinetics of binding and unbinding between DNA binding sites and dyes can be characterized by the rate (Equation 1) at which the occupation of a DNA binding site changes. New site-dye complexes form and old complexes disintegrate into empty sites and free dye molecules:

$$\frac{d[site \cdot dye]}{dt} = k_{\text{on}}[site][dye] - k_{\text{off}}[site \cdot dye] \tag{1}$$

Here [$site \cdot dye$] is the concentration of dye-bound DNA sites, [$site$] is the concentration of empty sites, [$dye$] is the concentration of free dye in solution and $k_{\text{on}}$ and $k_{\text{off}}$ are the on-rate and off-rate constants respectively. At equilibrium, Equation 1 is equal to zero. After rearrangement:

$$\frac{[site \cdot dye]_{\text{eq}}}{[dye]_{\text{eq}}[site]_{\text{eq}}} = \frac{k_{\text{on}}}{k_{\text{off}}} \equiv K_{\text{a}} \tag{2}$$

Here the last step follows from the definition of $K_{\text{a}}$, the association constant. Solving Equation 2 gives the concentrations of the components and allows the calculation of the on-rate and lifetime. For pre-equilibrium reactions, Equation 1 can be solved explicitly (Supplementary Information). Although to our knowledge reaction rate constants of SYTO-13 are not directly available, those of minor groove ligands exist, and allow similar conclusions to be drawn on the photoblinking nature of SYTO-13 over the time scale of our experiments [30].

*2.5.2 Simulations of DNA dye photoblinking*

Our simulations, written in Mathematica (Wolfram Research), model the behavior of fluorophores when experimental parameters are supplied to the code. A comparison between simulated and experimentally measured fluorescence pixel intensity values over time was made.

First, the analytical expression of the fluorescence emission intensity landscape was established: each emission event was modeled as a 2D Gaussian intensity profile on the camera pixel array of comparable PSF width to the real experimental data. For simplicity, in the first instance the intensity centroids of the fluorophores were selected as randomly chosen points from a straight-line approximation for the extent of a combed-out λ DNA molecule. Each fluorophore was assumed to undergo a number of bright/dark cycles before being permanently photobleached. The duration of each signal, the probability of each subsequent photoactive status being bright/dark, and the probability distribution of the duration of dark states (fit by an exponential distribution with an extended tail), were obtained from individually tracking a random selection of fluorophores.

The peak value of the profile was determined from the local laser excitation. Again, for simplicity, we assume a non-saturating regime such that the fluorescence emission intensity scales linearly with the amplitude of the exciting laser beam. We incorporate the fact that the excitation field itself has a Gaussian-shaped intensity profile across the lateral *xy* extent of the sample in the microscope focal plane. Parameters such as the fluorescence emission saturation point of the dye, and the correspondence between laser intensity and fluorophore brightness were obtained from the measured statistics of our experimental data. For the proportion of dye molecules in solution compared to those bound to DNA we used values from section *3.4.2*.

Finally, the digitization introduced by the camera itself was implemented into the simulation: the intensity distribution is divided up into distinct image frames and includes the dead time of the camera. The image was rasterized according to the physical pixel size of the camera. One sampling point was used per pixel: a point precisely in the center of the pixel. Lastly, the background noise was modeled as Gaussian white noise at comparable values to the experimental data, and added to each pixel intensity.

## 3. Results and Discussion

*3.1 Biochemical characterization of the DNA construct*

The λ DNA molecule is 48,502 base pairs (bp) in length. This makes agarose gel analysis for verification of addition of short sequences at the ends challenging since it does not produce visible shifts in the gel bands.

The gel image in Fig. 1c is a 0.7% agarose, SYBR-safe (Life Technologies) stained gel, run for 40 min at 100 V with a 1 kbp DNA ladder (New England Biolabs) in lane 1. In lane 2, a large amount of the stock concentration of λ DNA is stuck in the well due to its high molecular weight. As a consequence of this, the unavoidably more dilute λ DNA construct is too dilute to be visualized in the gel. As a result, microscopical controls for ensuring correct production of the λ DNA construct are preferable.

*3.2 Diffraction-limited 'pre-bleach' fluorescence images of DNA samples*

On initial illumination under standard diffraction-limited optical resolution with the laser we see continuous linear structures due to the expected high density labeling. We see a combination of combed-out sections of single molecules of DNA and DNA with a more globular appearance (Fig. 2a). Averaging over the combed out DNA molecules seen in four separate acquisitions, stretched strands of DNA with mean lengths 17.6 ± 2.5 μm (±s.d.) are seen, which is consistent with the expected length of double-stranded λ DNA of 16.3 μm [18]. Labeled strands of DNA that are not flowed out following washing steps in the sample incubation protocols are observed to have diameters of 1,700 ± 320 nm. Basic polymer physics modeling of DNA as a worm-like chain (for example see [31, 32] for typical applications of worm-like chain analysis to extended biopolymers) predicts an end-to-end length of $\sim(2L_cL_p)^{1/2}$, where $L_c$ is the total contour length (here ~18 μm) and $L_p$ is the persistence length (measured from earlier single-molecule optical-tweezers studies as ~50 nm [33]), which suggests an end-to-end length of 1.3-1.4 μm, broadly consistent within experimental error with our observations of globular DNA. In an assay where the DNA is surface-immobilized we expect small sections of the DNA molecule to be stretched between surface attachments, giving an experimental value marginally higher than the idealized non-tethered prediction, as we observe. In a 0.8 nM dilution of λ DNA incubated with YOYO-1 for 5 min and introduced to an immobilized DNA assay there are on average ~7 extended DNA strands and ~13 non-extended strands in a 256x256 pixel sub-array field of view (equating to an area 41x41 μm at the sample itself). The observed surface density was similar for SYTO-13.

Both the immobilized DNA assay (Supplementary Movie 1) and the tethered DNA assay (Supplementary Movie 2) produce fluorescently-labeled DNA strands. Since we have not utilized oxygen-scavenging systems in our assays we can visualize interesting phenomena when extended DNA strands break under prolonged laser illumination, due presumably to free-radical formation in the water solvent (Supplementary Movie 3, Fig. 2b), illustrating that the imaging system is capable of monitoring real-time changes to DNA topology at a single-molecule level. The immobilized DNA assay allows us to perform superresolution reconstructions, since the fluorophores remain in position even if damaged. The addition of the red organic dye Tex615 at one of the ends of the λ DNA construct allows us to use dual color channels to identify the point of attachment to the surface (Fig. 2b), demonstrating the imaging setup capability for performing dual color colocalization investigations at a single-molecule level (for example, see [34]). In preparation for work with a newly developed magneto-optical tweezers system combined with multicolor superresolution imaging [21], we can attach a magnetic bead to the DNA construct (Fig. 2d) via a biotin-streptavidin linkage.

*3.3 Post-bleach photoblinking images*

By maintaining laser illumination of the DNA construct we move from the regime in which we see continuous linear structures to a regime where we see sparse blinking behavior along the same DNA constructs. This is the requirement for superresolution imaging, and allows us to determine the position and intensity of individual molecules by fitting the point spread function model to the observed image. The blinking behavior we observe is consistent with our expectation from the model in section *2.5.2*, within the range of parameters quoted in the literature. This is evidence to support our claim that we observe photoblinking behavior rather than binding and unbinding behavior of the fluorescent dye.

Photoblinking can be achieved with both SYTO-13 and YOYO-1 (Fig. 3, Supplementary Movies 4 and 5) and produces data which can be reconstructed numerically (see section *2.5*).

*3.4 Superresolution data*

We analyzed the superresolution data from the three different point spread function (PSF) localization methods, and compared these data in light of the result from modeling of the stochastics of the photoactivity of the DNA dye molecules.

*3.4.1 Localization fitting*

Three software packages were used and compared for localization fitting the YOYO-1 and SYTO-13 labeled λ-DNA data, QuickPALM, rainSTORM and ADEMS code. To estimate true and false positive values we simulated data with similar noise and spot intensity to the experimental, creating a 54x54x1000 pixel array consisting typically of 10 spots. Table 2 summarizes the differences in performance between the three software packages.

Fig. 2 shows a fluorescence micrograph of YOYO-1 labeled DNA with ADEMS code reconstruction overlayed in red and cluster positions marked in black. Many more spots are found in the left half of the image as the laser intensity is brighter here. In rainSTORM, localization precision is estimated using the Thompson et al. formulation [35]. In ADEMS code, localization precision is estimated from the standard deviation of the centroid position of spots in multiple frames or from cluster analysis, by clustering spots which are very close together and measuring the mean distance between spots in a cluster (Fig. 2a). The localization precisions from each of these methods for the YOYO-1 and SYTO-13 dyes in shown in Table 3. SYTO-13 is generally more poorly localized as it is much dimmer but the different methods largely agree.

*3.4.2 Modeling the stochastic appearance of dye molecule images*

The binding constant of YOYO-1 to DNA, $K_a$, has reported values of $10^{10}$-$10^{12}$ M$^{-1}$[36-38]. Values two orders of magnitude lower have been reported [39], but the authors suggest that their low value was due to oversimplification of the equation used to model the binding kinetics. The ratio of bound dye to free dye is calculated to lie in the range $7\times10^5$:1-$7\times10^7$:1. Clearly the vast majority of dyes are bound to DNA in equilibrium. To evaluate the on-rate and the binding lifetime we take $k_{on} = 3.8\times10^5$M$^{-1}$ s$^{-1}$ [39]. It follows that in equilibrium only between $2.4\times10^{-3}$ and 0.24 dye molecules bind to a DNA molecule per second and that the average time a YOYO-1 dye molecule stays bound to DNA is between $2.6\times10^4$ s and $2.6\times10^6$ s, consistent with previous experimental measurements of the lifetime of YOYO-1 [40], which is in excess of the typical timescale of our experimental measurements (see Supplementary Information).

From solving Equation 1, we can determine the functional dependence of dye binding concentration with respect to time. The system reaches a 50% point of dye binding site occupancy in less than 100 ms of dye-DNA mixing (see Supplementary Information Fig 1).

An example simulation is shown in SI Movie 6. The initial image frames show a characteristic high density of photoactive dye molecules. The total integrated signal strength decays exponentially. After the initial image frames, individual fluorescence emission events are separated in space by a distance larger than the optical resolution limit, and so can be resolved, as we observe from the experimental data. Fig. 3a shows examples of fluorescence emission events of YOYO-1 bound to λ DNA which are reversibly photoblinking, with an example intensity vs. time trace for one dye molecule shown in Fig. 3b. From the analytical modeling of the binding kinetics in the previous section it is clear that stochastic photoactivity due to dynamic binding is negligible compared to photoblinking. If we approximate the dye 'off' time distribution as an exponential decay, the simulation displays essentially no photoactivity after ~5 seconds, incompatible with our observations of photoactivity after tens of seconds or more, but if we assume the 'off' time follows a non-exponential distribution with a more extended tail, the simulated fluorescence trajectories are in broad agreement with earlier findings [12].

## 4. Conclusions

By using the intrinsic reversible photoblinking of DNA binding dye molecules we can reconstruct details of the molecular structure of single DNA molecules *in vitro* with a localization precision which is less than the standard optical resolution limit in a diffraction-limited regime by almost an order of magnitude, both for a bright organic dye which intercalates, as well as for a dimmer dye which binds to the minor groove. The localization precision observed here of a few tens of nanometers could in principle be improved by reducing the time resolution of sampling, but decreasing the time resolution reduces the ability to resolve nearby fluorophores. We demonstrated our system's capability to capture images that can be super-resolved at video-rate sampling to show a promising future prospect for real-time monitoring of dynamic DNA topology. The successful application of an intercalating and a minor groove binding dye is important here since topological changes in DNA putatively affect the major and minor groove dimensions of double-stranded DNA differently; it may prove important to use a combination of both dye types to build up a representative picture of dynamic DNA topology which is not altered by the presence of the bound dye. Our future work will involve now combining this superresolution fluorescence imaging capability with single-molecule mechanical manipulation through magneto-optical tweezers we are developing which allow control of suitable micron sized beads attached to single DNA molecules, to ultimately permit superresolution fluorescence imaging of different topological states of single DNA molecules in real-time.

## Acknowledgements

Thanks to Christoph Baumann (University of York) and Stephen Cross (University of Bristol), for preliminary discussions in regards to DNA surface conjugation and blocking strategies, Peter O'Toole (University of York) for preliminary discussion regards superresolution imaging, and to Isabel Llorente-Garcia (UCL) and Quan Xue (University of Oxford) for preliminary contributions towards to the development of Matlab based superresolution fitting algorithms. This work was supported by funds from the Biological Physical Sciences Institute (BPSI) at the University of York.

**Figure and Tables**

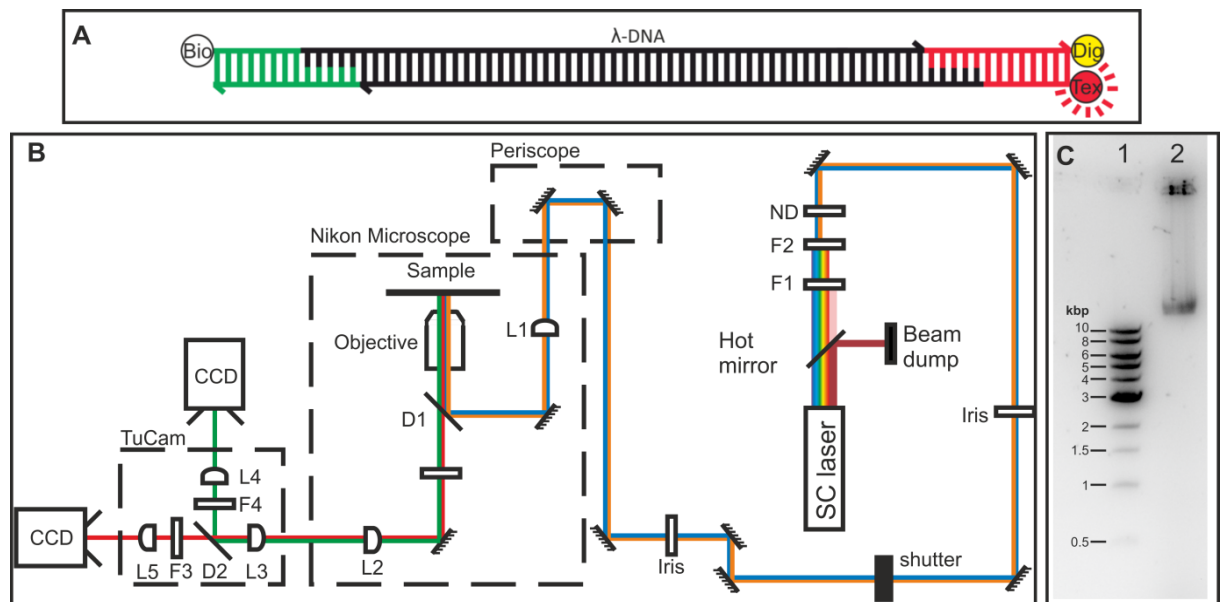

Fig. 1. DNA construct and bespoke optics. (A) Schematic of lambda-DNA construct (black) with the biotinylated synthetic DNA section (green) and digoxigenin-Tex615 section (red) indicated, 3' ends marked with a dash. (B) Schematic diagram of the multicolor superresolution fluorescence microscope. The infrared component of the laser emission is safely dumped using the hot mirror and filter F1. The resulting beam is then color-filtered and attenuated before entering the Nikon microscope body. On exit, it enters the TuCam commercial color splitter unit (Andor) and EMCCD camera system. (C) 0.7% Agarose gel showing a 1kbp ladder (lane 1) and the stock Lambda DNA (lane 2); a high percentage of the lambda DNA is stuck in the well.

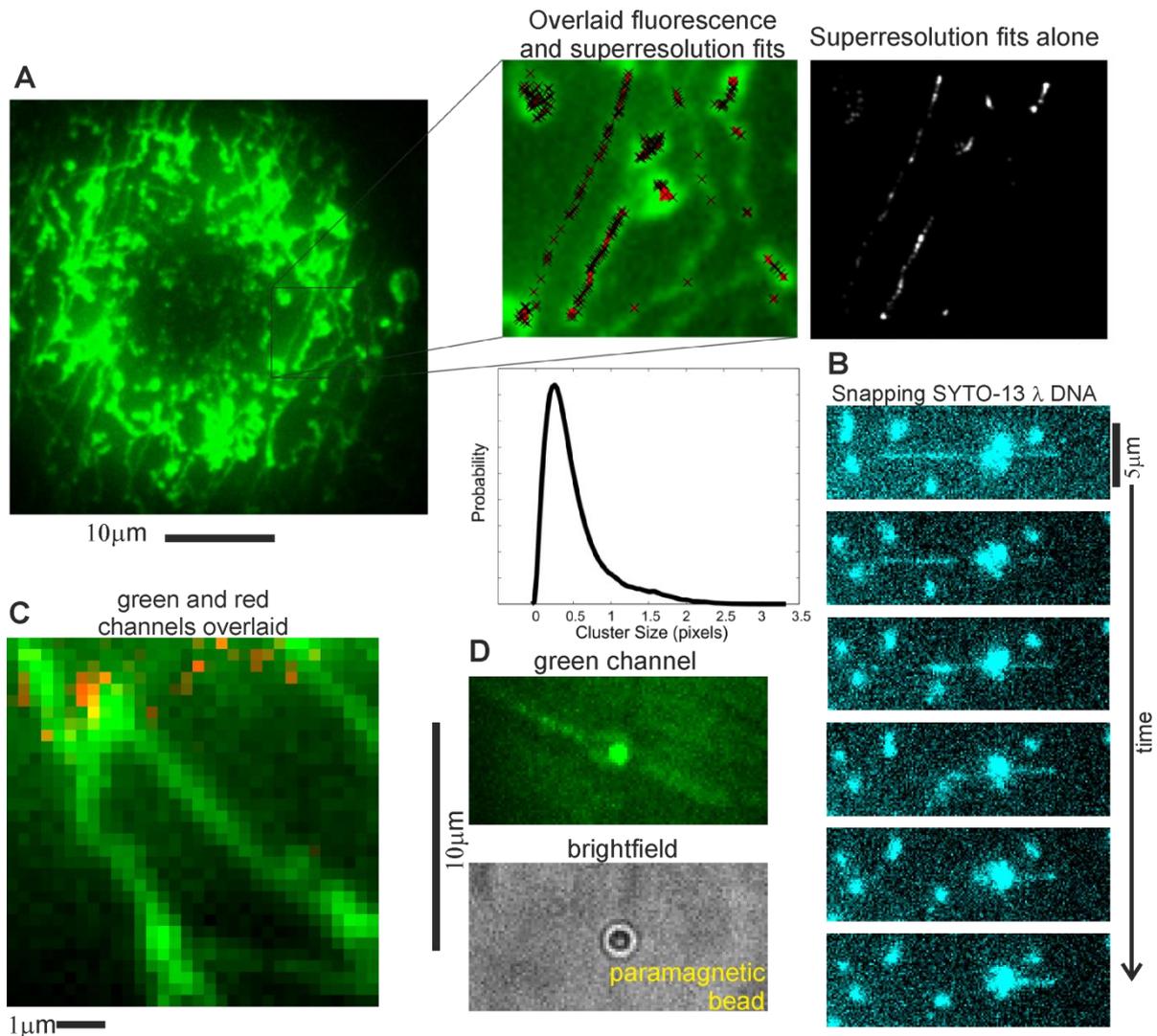

Fig. 2. Fluorescence imaging of DNA. (A) Fluorescence micrograph of YOYO-1 labeled λ DNA (green), expanded section shows ADEMS code reconstruction in red and clusters as black crosses (left panel) and the compiled superresolution fits alone (right panel). Also shown is the cluster size distribution of localizations from ADEMS code. (B) Sequential image frames of SYTO-13 labeled λ DNA initially combed straight and then snapped back following prolonged laser excitation. (C) Overlaid green channel (DNA) and red channel (DIG terminus of DNA construct). (D) Green channel and brightfield images of a paramagnetic bead attached to two strands of fluorescently labeled DNA.

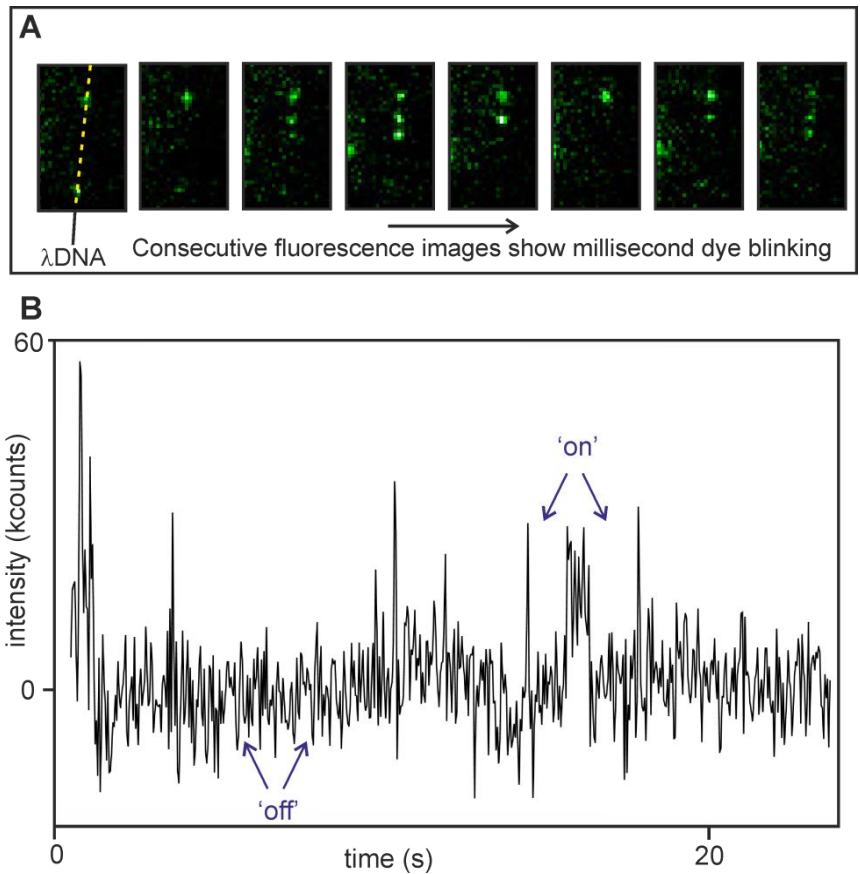

Fig. 3. Stochastic DNA-binding dye photoblinking. (A) Example of stochastic YOYO-1 photoblinking (green) from consecutive image frames, line of λ DNA marked (yellow). (B) Example Intensity vs time trace of a single YOYO-1 fluorophore bound to λ DNA, showing the integrated pixel intensity over each dye's PSF minus the local background intensity. We observed dark 'off' states lasting as up to ~10 s and bright 'on' states in the range of ~10-1000 ms.

| 1 | 5'-AGGTCGCCCCCGTTCGTTGAGTCA-digoxigenin-3' |
| 2 | 5'-Tex615-GACTCAACGAAC-3' |
| 3 | 5'-GGGCGGCGACCTGGACAGCAAGTTGGACAA-3' |
| 4 | 5'-biotin-TTGTCCAACTTGCTGTCC-3' |

Table 1: Sequences of synthetic oligonucleotides used to label λ DNA.

|  | **rainSTORM** | **ADEMS code** | **QuickPALM** |
|---|---|---|---|
| **Speed** | Parallelized for each reconstruction so very fast (<1s/1000 frames) | Only parallelized for batch processing (<1min/1000 frames) | Runs in java (<1s/1000 frames) |
| **Images** | Requires square images | Any image dimensions allowed | Any image dimensions allowed |
| **Precision characterization** | Thompson equation | Spot centroid distribution and cluster analysis | No precision analysis |
| **Percentage spots found on simulated data** | 78.3% | 67.3% | 48.5% |

Table 2: Summary of two different methods for localization fitting. The false positive rate we estimate as being low, in the range 0.1-1%.

|  | rainSTORM | ADEMS code – centroid distribution | ADEMS code – mean cluster distance |
| --- | --- | --- | --- |
| **YOYO-1** | 35 | 40 | 41 |
| **SYTO-13** | 67 | 62 | 90 |

Table 3: Summary of localization precision (rounded to nearest nm) for each method.

# Superresolution imaging of single DNA molecules using stochastic photoblinking of minor groove and intercalating dyes

Supplementary information

**Author names and affiliations:**

Helen Miller, Zhaokun Zhou, Adam J. M. Wollman, Mark C. Leake[*]

Biological Physical Sciences Institute (BPSI), Departments of Physics and Biology, University of York, York, YO10 5DD

[*] Corresponding author.

*E-mail address*: mark.leake@york.ac.uk (M. C. Leake)

# Contents

## S1 Equilibrium concentrations

Here we calculate the concentration of the various components involved in binding/unbinding kinetics in equilibrium.

For convenience, Equations 1 and 2 from the main text are reproduced here:

$$\frac{d[site \cdot dye]}{dt} = k_{on}[site][dye] - k_{off}[site \cdot dye] \tag{1}$$

$$\frac{[site \cdot dye]_{eq}}{[dye]_{eq}[site]_{eq}} = \frac{k_{on}}{k_{off}} \equiv K_a \tag{2}$$

From mass balance of the total amount of dye (free and bound) and DNA binding sites (empty and occupied):

$$[site] + [site \cdot dye] = [site]_0 \tag{S1}$$

$$[dye] + [site \cdot dye] = [dye]_0 \tag{S2}$$

where $[site]_0$ is $[site]$ at t=0 and $[dye]_0$ is $[dye]$ at t=0. Substituting (S1) and (S2) into Equation (2), we get:

$$\frac{[site \cdot dye]_{eq}}{([dye]_0 - [site \cdot dye]_{eq})([site]_0 - [site \cdot dye]_{eq})} = K_a \tag{S3}$$

This quadratic equation produces two solutions for $[site \cdot dye]_{eq}$, but only one has a sensible physical meaning:

$$[site \cdot dye]_{eq} = \frac{1}{2}([site]_0 + [dye]_0 + \frac{1}{K_a} \\ - \sqrt{\left([site]_0 + [dye]_0 + \frac{1}{K_a}\right)^2 - 4[site]_0[dye]_0}) \tag{S4}$$

Substitute in

$$[site]_0 = 1.59 \times 10^{-9} \times 48502 = 7.71182 \times 10^{-5} M$$

where $1.59 \times 10^{-9}$ is the concentration of DNA and 48,502 is the number of base pairs in λ DNA, and

$$[dye]_0 = 10^{-5} M \tag{S5}$$

and assuming $K_a = 10^{10} M^{-1}$ (taken as minimum reliable reported experimental value from previous investigations, see main text), we get:

$$[site \cdot dye]_{eq} = 9.9999851 \times 10^{-6} \text{M} \tag{S6}$$

therefore:

$$\frac{[site \cdot dye]_{eq}}{[dye]_{eq}} = \frac{[site \cdot dye]_{eq}}{[dye]_0 - [site \cdot dye]_{eq}} = 7 \times 10^5$$

Assuming that $K_a = 10^{12} \text{M}^{-1}$ (taken as maximum reliable reported experimental value from previous investigations, see main text), we get:

$$[site \cdot dye]_{eq} = 9.999999851 \times 10^{-6} \text{M} \tag{S7}$$

Therefore:

$$\frac{[site \cdot dye]_{eq}}{[dye]_{eq}} = 7 \times 10^7$$

## S2 Off-rate in equilibrium

To calculate the range of likely off-rate values, we first assume $K_a = 10^{10} \text{M}^{-1}$ as above, so the off-rate constant is given by:

$$k_{\text{off}} = \frac{k_{\text{on}}}{K_a} = \frac{3.8 \times 10^5}{10^{10}} = 3.8 \times 10^{-5} \text{s}^{-1} \tag{S8}$$

Note $k_{\text{off}}$ describes the intrinsic property of unbinding, not dependent on the concentration of the participating molecules. With the value of $[site \cdot dye]_{eq}$ from Equation S6, we get the total off-rate $R$:

$$R = k_{\text{off}}[site \cdot dye]_{eq} = 3.8 \times 10^{-5} \times 10.0 \times 10^{-6}$$
$$= 3.8 \times 10^{-10} \text{M s}^{-1} \tag{S9}$$

In a chamber of volume $v$, the number of unbinding events per each DNA molecule is:

$$\frac{[R]v}{[DNA]v} = \frac{3.8 \times 10^{-10}}{1.59 \times 10^{-9}} = 0.24 \tag{S10}$$

where $[DNA]$ is the concentration of DNA molecules in the flow chamber. Namely 0.24 dyes unbind from 1 DNA molecule per second. But, if we assume the higher experimental value of $K_a = 10^{12} \text{M}^{-1}$ [1], we get the off-rate constant:

$$k_{\text{off}} = \frac{k_{\text{on}}}{K_a} = \frac{3.8 \times 10^5}{10^{12}} = 3.8 \times 10^{-7} \text{s}^{-1} \tag{S11}$$

Or, the total off-rate $R$:

$$R = 3.8 \times 10^{-7} \times 10.0 \times 10^{-6} = 3.8 \times 10^{-12} \text{M s}^{-1} \tag{S12}$$

And the number of unbinding events per DNA per second is $2.4 \times 10^{-3}$.

**S3 Time to approach equilibrium**

To find out how long it takes to reach 50% of equilibrium occupancy of dye binding sites when DNA is mixed with dye, we re-write Equation 1 as:

$$\frac{dx}{dt} = k_{on}([site]_0 - x)([dye]_0 - x) - k_{off}x$$

where $x \equiv [site \cdot dye]$, and solve:

$$\frac{dx}{k_{on}([site]_0 - x)([dye]_0 - x) - k_{off}x} = dt$$

Resulting in the following curve:

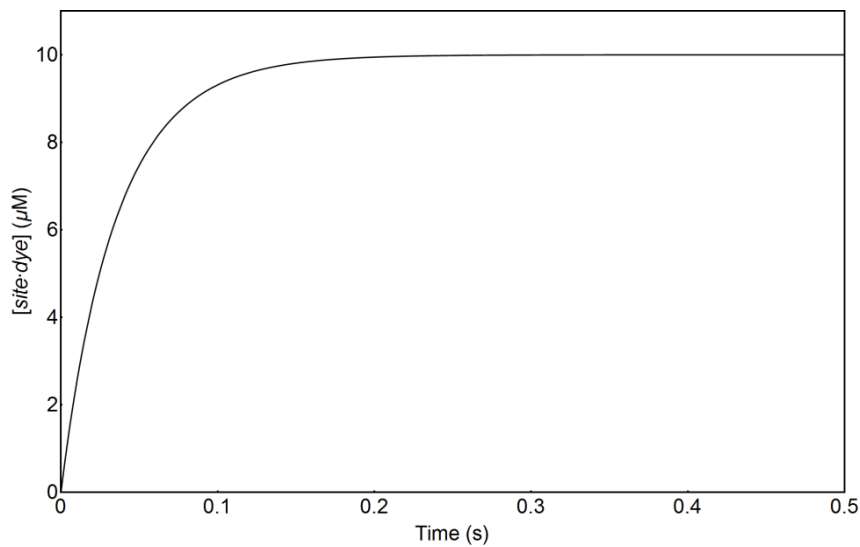

Supplementary Fig. 1. Kinetics of dye binding to DNA.

We find that the 50% binding equilibrium point is reached after ~30 ms.